\documentclass[aps,preprint,nofootinbib,preprintnumbers]{revtex4-1}
\usepackage{graphicx}
\usepackage{amsmath,amssymb,bm}
\usepackage[mathscr]{eucal}
\usepackage[colorlinks,linkcolor=blue,citecolor=blue]{hyperref}

\newif\ifarxiv
\arxivfalse

\begin		{document}
\def\st    {\begin{equation}}
\def\stp    {\end{equation}}

\def\d		{d}
\def\Del		{\nabla}

\def\half	{{\textstyle \frac 12}}

\def\Arxiv      #1 [#2]{\href{http://arxiv.org/abs/#1}{{\tt arXiv:#1 [#2]}}\,}

\def\half{{\textstyle\frac{1}{2}}}

\def\CA{{a}}
\def\CJ{{j}}

\definecolor{Blueberry}{rgb}{0.25,0,0.65}
\definecolor{Strawberry}{rgb}{0.65,0,0.25}

\newcommand{\be}{\begin{equation}}
\newcommand{\ee}{\end{equation}}
\newcommand{\bea}{\begin{eqnarray}}
\newcommand{\eea}{\end{eqnarray}}
\newcommand{\bwt}{\begin{widetext}}
\newcommand{\ewt}{\end{widetext}}

\newcommand{\bi}{\begin{itemize}}
\newcommand{\ei}{\end{itemize}}
\newcommand{\ben}{\begin{enumerate}}
\newcommand{\een}{\end{enumerate}}
\newcommand{\bca}{\begin{cases}}
\newcommand{\eca}{\end{cases}}
\newcommand{\bln}{\begin{align}}
\newcommand{\eln}{\end{align}}
\newcommand{\bst}{\begin{split}}
\newcommand{\est}{\end{split}}

\newcommand\Lam{\Lambda}

\newcommand\Ga{{\ensuremath{{\Gamma}}}}

\newcommand\De{{\ensuremath{{\Delta}}}}

\newcommand\nab{{\nabla}}

\def\th{{\theta}}

\newcommand\ha{{\half}}

\def\le{\left}
\def\ri{\right}

\newcommand\sT{{\mathcal T}}

\newcommand\p{\ensuremath{\partial}}

\newcommand\vev[1]{{\ensuremath{\left\langle{#1}\right\rangle}}}

\title
    {
   Holographic Vortex Liquids and Superfluid Turbulence
    }
\author{Allan~Adams}
\author{Paul~M.~Chesler}
\author{Hong~Liu}

\affiliation
    {
Department of Physics, 
MIT, 
Cambridge, MA 02139, USA 
    }

\date{\today}

\begin{abstract}

{Superfluid turbulence, often referred to as quantum turbulence, is a fascinating phenomenon} for which a satisfactory theoretical framework is lacking.  
Holographic duality provides a systematic new approach to studying quantum turbulence by {mapping} the dynamics of 
{certain} 
quantum {theories onto} the dynamics of classical gravity.  
{We use this gravitational description to numerically construct turbulent flows 
in a holographic superfluid in two spatial dimensions.
We find that the superfluid kinetic energy spectrum obeys the Kolmogorov $-5/3$ scaling law,  as it does for turbulent flows in normal fluids. }
We trace this scaling to a direct energy cascade by injecting energy at long wavelengths and watching it flow to a short-distance scale set by the vortex core size, where dissipation by {vortex annihilation and vortex drag} becomes efficient.  
This is in sharp contrast with the inverse energy cascade of normal fluid turbulence in two dimensions.
We also demonstrate that the microscopic dissipation spectrum has a simple geometric interpretation.

\end{abstract}

\preprint{MIT-CTP-4419}

\pacs{}

\maketitle

\section{Introduction}

{
Superfluid turbulence is a fascinating non-equilibrium phenomenon dominated by the dynamics of quantized 
vortices~\cite{Feynman,Viven1,Viven2} (for recent reviews see~\cite{VivenR,Paoletti2011,Tsubota2012}).
In contrast to normal fluids, which are well-described in the turbulent regime by 
dissipative hydrodynamics,  
superfluids exit the hydrodynamic regime when quantized vortices are present.
Considerable insight has nonetheless been gained from phenomenological models of vortex dynamics, 
particularly thanks to powerful numerical simulations which play a central role in any discussion of turbulence.
Nonetheless, these phenomenological models have significant limitations and shortcomings, and a 
satisfactory theoretical framework describing superfluid dynamics remains lacking.
An {\it ab initio} study would be greatly desirable. 
}

In this paper we initiate a 
study of
superfluid 
{turbulence} 
using holographic duality and report 
new results in two spatial dimensions.  Holographic duality equates certain systems of quantum matter without gravity to 
classical
gravitational
systems in a curved spacetime with one additional 
{spatial dimension (\cite{Maldacena:1997re,Gubser:1998bc,Witten:1998qj}, see {\em e.g.} \cite{Hartnoll:2009sz,Herzog:2009xv,McGreevy:2009xe,CasalderreySolana:2011us,Adams:2012th}\ for recent reviews).
Holographic duality provides a complete description --- valid at all scales ---} of a strongly interacting quantum many-body system in terms of  a classical gravitational {system.}  
It thus allows a first-principles study of the superfluid, including turbulent flows, by using the corresponding gravity description of the superfluid phase. 
Furthermore, {the} gravity description provides a new geometric reorganization of turbulent dynamics.  {For example,  dissipation in the gravitational description can be understood
in terms of excitations falling through a black hole event horizon. } This provides a direct measure of the rate of energy dissipation and its spectrum.

{We focus on  ``non-counterflow'' superfluid turbulence, which has been the subject of considerable experimental and numerical study during the last decade} (see for 
example~\cite{Maurer98,smith93,stalp99,Henn2009PRL,Seman2011,nore97,araki02,kobayashi05,parker05}). 
Among the most significant results of these studies is the observation of Kolmogorov's $-{5 \over 3}$ scaling law in the kinetic energy spectrum, which suggest that quantum and classical turbulence {may} share certain statistical properties characterized by the Kolmogorov law despite the fundamental differences between an ordinary fluid and a superfluid. 
In classical turbulence {this} scaling behavior can be understood as a consequence of an energy cascade {in which} the injected energy is passed from one scale to another without 
substantial loss. Whether the observed quantum turbulence {admits a similar cascade picture, and if so whether the cascade drives energy to long or to short wavelengths,} remain important open questions.

The comparison between classical and quantum turbulence is particularly sharp in two spatial dimensions.
In two spatial dimensions, the enstrophy density (the square of the vorticity) of a classical fluid is conserved.  This implies that colliding vortices of opposite vorticity cannot annihilate. 
In contrast, vortices of similar vorticity can merge and produce larger and larger vortices.
Indeed, Kraichnan \cite{Kraichnan:1967aa} argued that the conservation of enstrophy in non-relativistic turbulent flows implies that energy must be transported from the UV to the IR in an inverse cascade.   An inverse cascade and enstrophy conservation have recently been demonstrated in relativistic conformal fluids in two spatial dimensions as well \cite{Carrasco:2012nf}.  
This behavior stands in stark contrast to a superfluid, where enstrophy is not conserved, vortex annihilation is allowed and vortex merging is energetically suppressed.   
Moreover, even in regimes in which vortex annihilation is negligible so that enstrophy is effectively conserved, the energetic arguments used by Kraichnan do not apply to the non-hydrodynamic vortex liquid.
Simply put, the mechanics of quantized vortices are different than the mechanics of vortices in normal liquids.
We therefore have no a priori expectation for the direction of a turbulent cascade in a two dimensional superfluid.  Indeed, several recent numerical studies 
of the phenomenological Gross-Pitaevskii equation (with dissipation put in by hand) observed Kolmogorov scaling but came to conflicting conclusions regarding the direction of cascade~\cite{horng,numasato1,numasato2,reeves}.

In this paper we numerically construct turbulent flows in a holographic superfluid in two spatial dimensions by solving {the} equations of motion of the gravity 
dual.  We focus on flows in which the net vorticity is zero.  
When the flow approaches a turbulent quasi-steady-state, the system exhibits a scaling regime which obeys the Kolmogorov law. By driving the system with a long wavelength source in the scaling regime and examining the resulting energy flux through the black hole horizon, we demonstrate that the system exhibits a direct energy cascade, i.e.  the injected energy flows through an inertial range to a smaller length scale {of the order of a vortex core size,} where it gets dissipated through  vortex drag {and vortex annihilation}. 
{These results are derived from first-principles,} 
with no {phenomenological} assumptions made on the vortex dynamics of the superfluid or dissipation mechanism.

\section{The Holographic Set-up} \label{sec:set}%
We begin by setting up the gravity description of a holographic superfluid.  {We would like to construct} a quantum field theory in two spatial dimensions with a complex scalar operator, $\psi(x)$, carrying charge $q$ under a global $U(1)$ symmetry.  Let $j^{\mu}(x)$ denote the conserved current operator of this global $U(1)$ {symmetry.}  
To induce a superfluid condensate for $\psi$, we will turn on a chemical potential $\mu$ for the $U(1)$ charge.  
{For sufficiently large $\mu$, we expect $\psi$ to develop a nonzero expectation value $\vev{\psi} \neq 0$ when the temperature falls below a critical temperature $T_c$, spontaneously breaking the global $U(1)$ symmetry and driving the system into a superfluid phase. }

{A simple holographic system with this structure begins with a classical field theory living in an asymptotically anti-de Sitter spacetime with 3 spatial dimensions (AdS$_4$).
Under the standard holographic dictionary, the conserved current $j^\mu(x)$ is mapped to a dynamical $U(1)$ gauge field $A_M(x,z)$ in the gravitational bulk, while the scalar operator $\psi(x)$ is mapped to a bulk scalar field $\Phi(x,z)$ carrying charge $q$ under the gauge field $A_M$.  {Note $z$ is the radial coordinate of AdS$_4$.}%
\footnote
  {
  We label boundary indices by $\mu, \nu, \cdots$, and bulk indices by $M, N, \cdots$, with $A_M = (A_\mu,A_z)$.
  }
Placing the system at nonzero temperature corresponds to adding to the bulk spacetime a black hole whose horizon is a two-dimensional plane extended in boundary spatial directions. 
Adding a chemical potential corresponds to imposing a boundary condition on the bulk gauge field $A_{t}=\mu$ at the boundary of $AdS_{4}$.%
\footnote
  {
  {Notably this gravitational system is dual to a conformal field theory; however, conformal symmetry 
  is broken by both the chemical potential and the temperature, so the conformal symmetry will plan no role in what follows. }
  }
As found in~\cite{Gubser:2008px,Hartnoll:2008vx}, if the charge $q$ and scaling dimension $\De$ of $\psi$ lie in certain range, taking $\mu$ sufficiently large drives the bulk scalar field $\Phi$ to condense through the Higgs mechanism, so that the black hole develops scalar ``hair'' of $\Phi$ outside the horizon.}

There are many examples of {quantum} theories with a low-temperature superfluid phase which 
admit such a {gravitational} description~\cite{Denef:2009tp,Gubser:2009qm}. 
A universal bulk description for them is an Abelian Higgs model of $A_M$ and $\Phi$ coupled to the Einstein gravity, with different systems having different charge $q$ and potentials for $\Phi$. 
For definiteness we will choose a quadratic potential with a mass for $\Phi$ correspond to $\psi$ having scaling dimension $\De = 2$ as in~\cite{Hartnoll:2008vx}. We will work in the probe limit of~\cite{Hartnoll:2008vx}, which applies when the charge $q$ of $\Phi$ is large.  
In this limit  the {gravitational system is approximated by 
an Abelian Higgs {model} of $A_M$ and $\Phi$ in a Schwarzschild black hole 
geometry, with the backreaction of $A_M$ and $\Phi$ on the geometry neglected.}
See Appendix~\ref{app:A} for details on the black hole metric and the bulk action we use. 
The probe limit is {appropriate for studying} superfluid turbulence in a regime where the charged components 
of the fluid (both normal and superfluid) do not interact with the uncharged component of the superfluid.
 
A superfluid state 
generically has gapped vortex excitations, which will play an important role in our discussion below. 
Around a vortex, the fluid circulation is quantized. 
Introducing the (un-normalized) superfluid velocity 
\be\label{eq:suvo}
\bm u \equiv \bm {\mathcal J}/|\langle \psi \rangle|^2, \qquad 
\bm {\mathcal J} \equiv \frac{i}{2 }
\left[ \langle \psi^* \rangle \Del \langle \psi \rangle - \langle \psi \rangle \Del  \langle \psi \rangle ^*\right]
\ee
the winding number $W$ of a vortex is determined by 
\be 
W = \frac{1}{2 \pi} \oint_{\Gamma} d \bm x \cdot \bm u \ ,
\ee
where the path $\Gamma$ encloses a single vortex and is oriented
counterclockwise. {Here we {have} used bold-faced symbols to denote vectors along boundary spatial directions; $\bm x \equiv \{x_1,x_2\}$ with $x_i$ the two spatial directions
and $\nab = \{\frac{\partial}{\partial x_1}, \frac{\partial}{\partial x_2} \}$}.  
{A vortex {maps into} the gravitational bulk as a flux tube along the AdS} radial direction, stretching from the boundary, where they have a characteristic size $1/\mu$, to the horizon. Inside the flux tube the condensate goes to zero, effectively punching a hole through the bulk scalar condensate of characteristic coordinate size $1/\mu$. 
Explicit gravity 
{solutions}
corresponding to a static vortex of arbitrary winding number were  previously  constructed numerically in \cite{Albash:2009iq, Montull:2009fe,Keranen:2009re}. 
\begin{figure}[h]
\begin{center}
\includegraphics[scale=0.4]{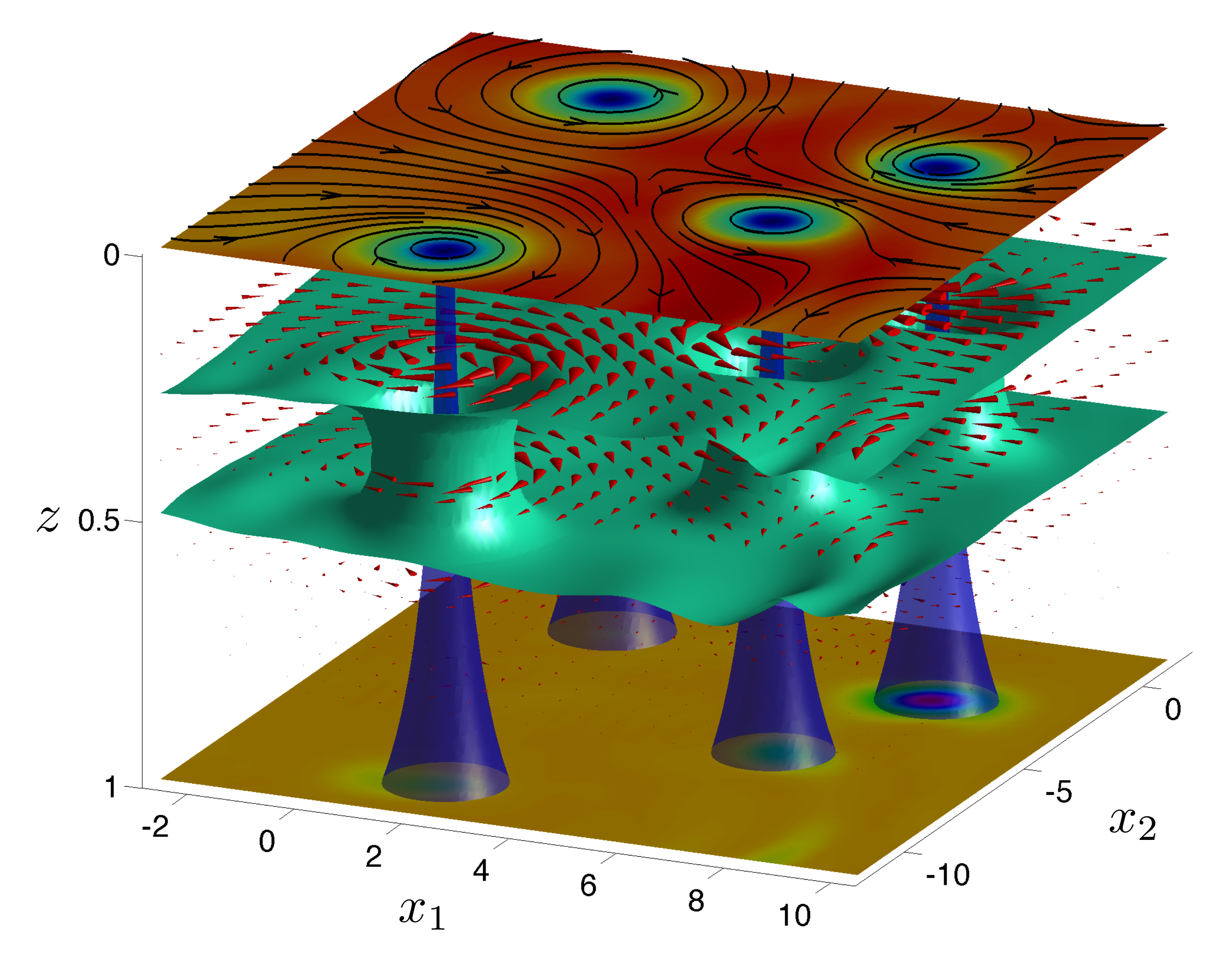}
\end{center}
\vskip -1.cm
\caption{{Holographic} description of a superfluid with vortices. 
The vertical axis is the radial direction $z$ {of AdS$_4$}.
 The planes $z=0$ and $z=1$ are the boundary {of AdS$_4$} and the black hole horizon respectively. 
 The green surface is a surface of constant bulk charge density, with the region between the two slices defining a ``slab'' of condensate where most bulk charges reside~(see Appendix~\ref{app:A} for details on the distribution of charge density in the bulk). 
{The slab screens excitations from falling into the horizon.} This can be seen from the vector field in the plot which gives energy flux $(-\tau^{x}_{0},-\tau^{y}_{0},-\tau^{z}_{0})$ of~\eqref{eq:deft};  the vector field (whose length represents its amplitude) vanishes very quickly below the slab.
 The vortices, with energy flux circulating around them, punch holes through this screening slab, providing avenues for {excitations} to fall into the black hole. 
  The surface $z = 0$ also shows the condensate on the boundary (with blue color representing zero condensate), superposed with flow lines of the superfluid velocity~\eqref{eq:suvo}. The flux tubes show a surface of {constant}  $|\Phi|^2/z^4$, which coincides with the boundary condensate at $z=0$. The $z = 1$ surface also shows the flux of energy through the {horizon.  Note that the energy flux is only significant (red and green) in the wake of the moving vortices.}
}
\label{fig:candy}
\end{figure}

{The gravity dual thus provides a first-principles description of superfluid flows involving vortices.  In addition, it also provides effective tools to describe and visualize  
dissipation in the system.} 
{Consider turning on a perturbation of $\langle j^\mu \rangle$ in the boundary theory, which on the gravity side 
corresponds to turning on a perturbation of $A^M$ near the boundary.}  Above $T_c$, the disturbance quickly falls into the black hole, 
{corresponding in the quantum theory to the perturbation in the current $\langle j^\mu \rangle$ quickly dissipating into heat.}
Below $T_c$, however, the scalar condensate essentially ``screens'' the black hole from boundary. As a result a $U(1)$ disturbance cannot reach the horizon to get dissipated and the perturbation
in the current $\langle j^\mu \rangle$ persists. This is the bulk realization of the non-dissipative nature of a superfluid. Now suppose
the superfluid has some vortices. Since the flux tube corresponding to a vortex punches a hole of zero condensate through the bulk scalar condensate, it provides an avenue for perturbations near the boundary to pass unimpeded to the horizon.  This implies that in the boundary system vortices could dissipate modes of wavelengths smaller than typical vortex size, but not those with larger wavelengths. See Fig.~\ref{fig:candy}.
{As we will see below, this heuristic picture efficiently encodes {much} of the physics of the system.}

The above discussion can be quantified by ``measuring'' the energy flux through the black hole horizon. 
In the probe limit we are working with, such {a} flux is particularly simple to define. Let $\mathcal T^{M}_{\ N}$ denote the stress tensor of $A_M$ and $\Phi$ in the bulk. 
Covariant conservation of $\mathcal T^{M}_{\ N}$ implies that the following bulk tensor 
\be \label{eq:deft}
\tau^{M}{_\mu} \equiv \sqrt{-g} \mathcal T^{M}{_\mu}
\ee
is conserved 
\begin{equation} \label{eq:cons}
\partial_\mu \tau^{\mu}_{\ \nu} = -\partial_z \tau^{z}_{\ \nu}\ 
\end{equation}
where $g$ is the determinant of the bulk metric. 
Equation~\eqref{eq:cons} has the simple interpretation that the non-conservation of $\tau^{\mu}_{\ \nu}$ along the boundary directions 
is equal to the flux $\tau^{z}_{\ \nu}$ along the radial direction. Of  particular interests is  the (positive) flux of energy through the horizon
\begin{equation} \label{eq:fl}
Q_{\rm horizon}(t) \equiv -\int d^2 x \, \tau^{z}_{\ t}(t,\bm x,z) \big |_{\rm horizon} \ .
\end{equation}
The energy that flows {across} the horizon into the black hole
is irreversibly lost and should be thought of as energy lost to heat.
See Appendix~\ref{app:A} for the explicit expression of $\mathcal T^{M}_{\ N}$ as well as 
other properties of $\tau^M{_\mu}$.  {As Fig.~\ref{fig:candy} suggests and as we discuss in 
greater detail below, the energy flux is localized at the locations of flux tubes in the bulk and hence
vortices in the superfluid.}

\section{Turbulent Flows and Kolmogorov scaling} \label{sec:tub}

{We now describe turbulent flows in the superfluid which we constructed by numerically solving the bulk equations of motion for a variety of initial 
conditions.\footnote{Images and videos from these simulations are available at \href{http://turbulent.lns.mit.edu/Superfluid}{http://turbulent.lns.mit.edu/Superfluid}.}
Working in units in which the temperature is $T = 3/(4 \pi)$, we set the chemical potential to be $\mu = 6$  and work in a $100\times100$ periodic box.  Since different  initial conditions lead to qualitatively similar late-time behaviors,  we focus for definiteness on a typical example.}

\begin{figure}[t!]
\includegraphics[scale = 0.45]{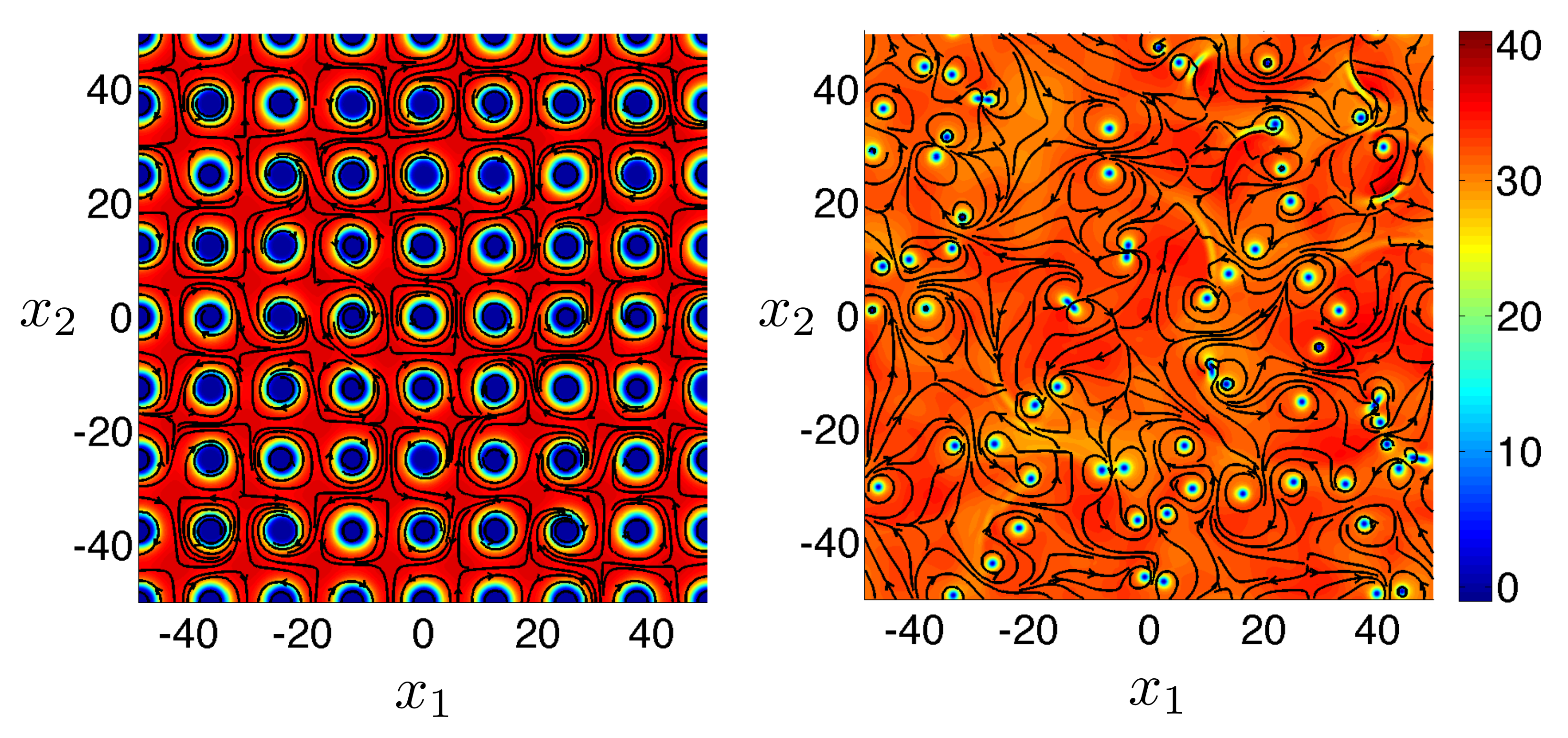} 
\vskip -0.05in
\caption{The superfluid condensate $| \langle \psi(t,\bm x) \rangle |^2$
at time $t=0$ (left) and $t = 300$ (right), with flow lines of the superfluid current $\bm {\mathcal J}(t,\bm x)$ (defined in~\eqref{eq:suvo}) superimposed.
The superfluid current circulates around the core of each vortex, where the condensate vanishes.   The winding number $\pm 6$ vortices (left) are much larger than the 
winding number $\pm 1$ vortices (right). 
The yellow arcs seen in the right figure are waves produced by the annihilation of vortex pairs. 
\label{fig:condensate}
} 
\end{figure}

We take as our initial condition a square periodic lattice
of winding number $W = \pm 6$ vortices, with winding number alternating 
at each lattice site and with lattice spacing  $b = 100/8$.%
\footnote
  {
  As we discuss below, the winding number $W = \pm 6$ vortices {rapidly} decay into 
  six winding number $W = \pm 1$ vortices.  
  Therefore, by adjusting the lattice constant {and initial winding number} these initial conditions allow us to control the initial
  density of {winding number $W = \pm1$ vortices.}
  }
 See the left panel of Fig.~\ref{fig:condensate}.
We evolve the system for a total period of time $\Delta t = 600$. 
The evolution of the system can be roughly  divided into three stages: (i) {a} homogenization 
regime {($t < 160$)}; (ii) {a} scaling regime {($160 < t < 500$)}; and (iii) {a} relaxation regime {($t > 500$)}. Turbulent behavior including the Kolmogorov scaling in the kinetic energy is observed in the scaling regime.

In the homogenization regime the system evolves from an ordered, inhomogeneous  initial state shown in the left panel of Fig.~\ref{fig:condensate} to a chaotic, quasi-homogeneous state shown in the right panel of Fig.~\ref{fig:condensate}. 
This regime includes the explosive decay of our initial $W=\pm6$ vortices to six smaller $W = \pm 1$ vortices.  This decay
{generically occurs without any intermediate stage of winding number $6>|W| > 1$ vortices.}
As time passes, vortices of opposite winding number $W = \pm1$ collide and annihilate. 
Since vortices in the superfluid are gapped excitations with the gap scaling like $W^2$,
the merging of vortices is heavily suppressed energetically and indeed {has never been} observed in our simulations.

{
The scaling regime begins to set in around time $t = 160$, which roughly 
corresponds to the time in which the $W = \pm 1$ vortex annihilation rate begins to dramatically slow down.}
The system is now turbulent, characterized by a random, yet homogeneous (at large scales)
distribution of a large number of vortices and anti-vortices of unit winding (see the right panel of Fig.~\ref{fig:condensate}). 
The motion of the vortices is highly irregular, with the location and velocity of the vortex cores
at any given time extremely sensitive to initial conditions. 

The defining feature of the scaling regime is that 
the system exhibits the Kolmogorov's $-5/3$ scaling law. A particularly nice observable to see the scaling behavior is the ``kinetic energy" density
\begin{equation}
\label{eq:landauginz}
\mathcal E_{\rm kin}(t, \bm x) \equiv \frac{1}{2 } \bm {\mathcal V}^*(t,\bm x) \cdot \bm {\mathcal V}(t,\bm x),
\end{equation}
where $\bm {\mathcal V} = \langle \psi \rangle \bm u$.  Introducing a spatial Fourier transform, the total ``kinetic energy'' can
be written as an integral over momentum
\be
E_{\rm kin}(t) = \int d^2 x \, \mathcal E_{\rm kin}(t, \bm x) = \int_{0}^{\infty} dk \, \epsilon_{\rm kin}(t,k)
\ee
where 
\be \label{eq:ens}
\epsilon_{\rm kin}(t,k) = \frac{1}{2 }\int_0^{2 \pi} d\theta \, k \bm {\mathcal V}^*(t,\bm k) \cdot  \bm {\mathcal V}(t,\bm k)
\ee
with the $\theta$ integral summing over directions of $\bm k$. Note that while $\mathcal E_{\rm kin}(t, \bm x) $ and the associated $\epsilon_{\rm kin}(t,k)$ {are well-defined observables} for any quantum many-body {system, their interpretation} as kinetic energy density in coordinate and momentum space  is at most heuristic, as in a strongly interacting quantum system there is really no unambiguous 
way to define the kinetic energy.
{We note, however, that up to constants our expression for the superfluid kinetic energy (\ref{eq:landauginz}) agrees with the usual expression for the superfluid kinetic energy in the non-relativistic hydrodynamic limit.}

\begin{figure}[h!]
\includegraphics[scale = 0.45]{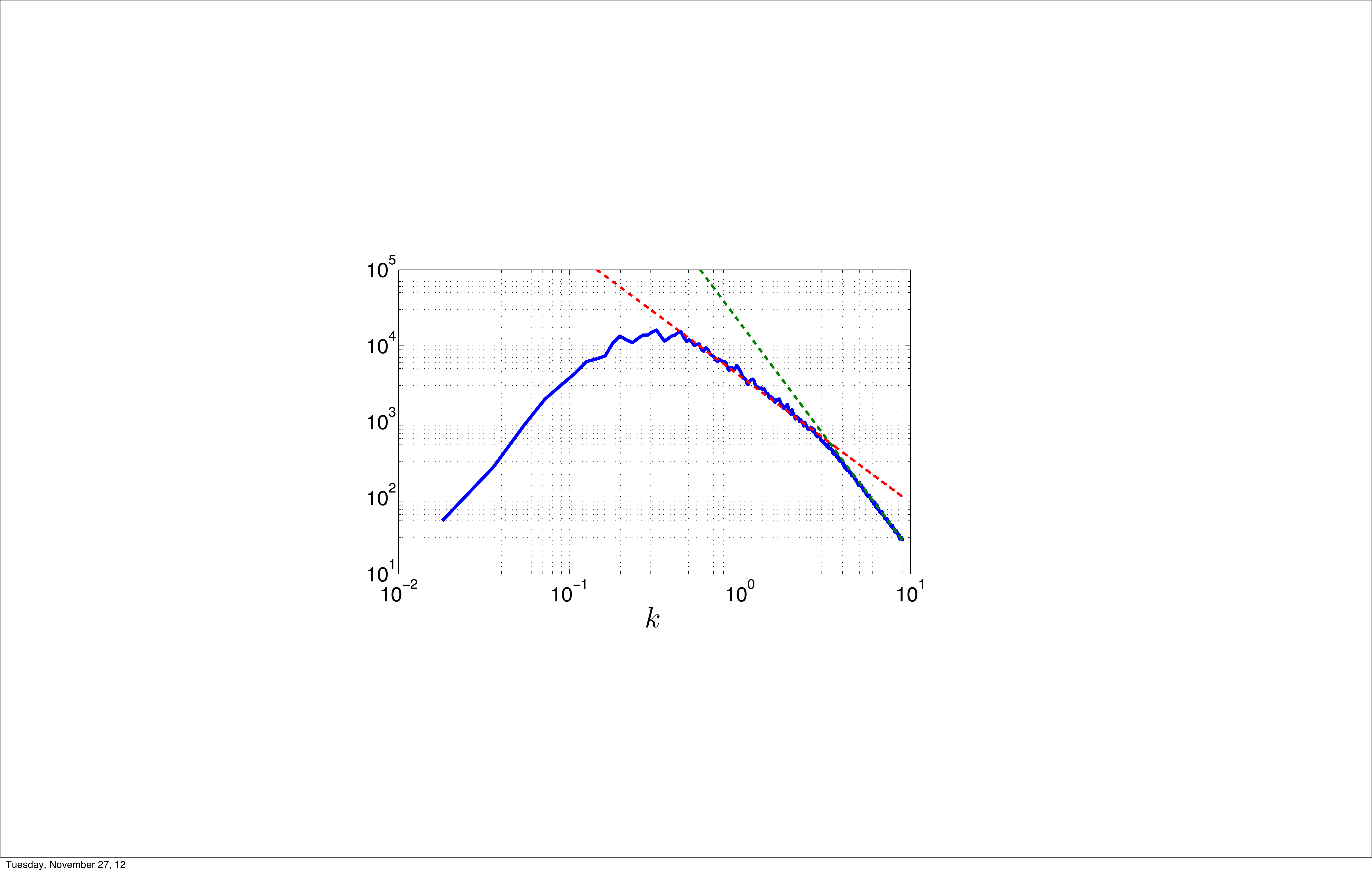}
\vskip -0.05in
\caption{The energy spectrum $\epsilon_{\rm kin}(t,k)$ at time $ t = 300$.
The red dashed line is the Kolmorgorov scaling $k^{-5/3}$ and the green dashed line is $k^{-3}$.  
\label{fig:powerspectrum}
} 
\end{figure}

In Fig.~\ref{fig:powerspectrum}, we plot $\epsilon_{\rm kin}(t,k)$ at the same time shown {as the right panel of} Fig.~\ref{fig:condensate}, $t = 300$.  
Also included in Fig.~\ref{fig:powerspectrum} is the {curve $k^{-5/3}$}.  Remarkably, the 
energy spectrum obeys the $k^{-5/3}$ scaling in the {interval $k \in \{0.4,3\}$} which translates into 
length {scales $(2, 16)$} (recall our box size is $100$). The average vortex spacing at this time is about $10$, falling in the middle of the scaling region.

The $k^{-5/3}$ scaling persists until the end of our simulation $t=600$, but for $t > 500$, the scaling behavior becomes less and less {sharp.} 
{By the time $t = 500$, due to vortex annihilation, the number of $W = \pm 1$ vortices has decreased by $O(10)$ from its maximum value.}  
Notably, for initial data whose evolution does not generate any vortices, we do not find any universal scaling behavior of $\epsilon_{\rm kin}$.  
Evidently, the scaling behavior $\epsilon_{\rm kin} \sim k^{-5/3}$ crucially depends on having a homogenous 
vortex liquid.

It is useful to recall Kolmogorov's logic for the derivation of the $\epsilon_{\rm kin} \sim k^{-5/3}$
scaling.  Kolmogorov assumed the existence of an \textit{inertial range} $k \in  (\Lambda_{-} , \Lambda_+)$ 
where the energy spectrum (per unit mass) only depends on the scale $k$ and the mean rate of energy dissipation per unit mass $\varepsilon $.  
With these assumptions, non-relativistic dimensional analysis then yields $\epsilon_{\rm kin} \sim \varepsilon^{2/3} k^{-5/3}$. In our system we have $\Lam_+ \approx 3$ and $\Lam_- \approx 0.4$.

{Fig.~\ref{fig:powerspectrum} also shows a power law $\epsilon_{\rm kin} \sim k^{-3}$ for $k > 3$. This behavior appears as soon as the initial winding number $6$ vortices decay and persists until the end of our simulation. 
This scaling arrises from the short-distance behavior of $\bm{ \mathcal V} ({\bm x}) $ near vortex cores, and thus reflects 
single vortex physics and not collective physics or turbulence.  
In particular, near a winding number $\pm 1$ vortex core the superfluid velocity scales like $\bm u \sim \hat \theta /d$ where $d$ is the distance to the core and $\hat \theta$ the angular direction around the core.
Similarly, the condensate vanishes like $\langle \psi \rangle \sim d$.  It follows that near a vortex core $\bm {\mathcal V} \sim \hat \theta $ so $\bm {\mathcal V}$ is not continuous.   This discontinuity implies that in Fourier space $\bm {\mathcal  V} \sim k^{-2}$ at large $k$ and therefore that $\epsilon_{\rm kin} \sim k^{-3}$ at large $k$.  Evidently, the transition from the collective physics responsible for the $\epsilon_{\rm kin} \sim k^{-5/3}$
scaling to the single vortex physics responsible for the $\epsilon_{\rm kin} \sim k^{-3}$ scaling occurs around $k \approx 3$.}

{We note that in a well defined sense our system is non-relativistic during the scaling regime.  One can define a normalized 3-velocity in terms of the expectation value of the electromagnetic current: 
$v^\mu \equiv \langle j^\mu \rangle /\sqrt{-\eta_{\alpha \beta} \langle j^\alpha \rangle \langle j^\beta \rangle}$.   During the scaling regime the average value of $|\bm v|$ is never greater than $0.05$.
We note furthermore that if $\bm v$ is used in place of $\bm u$ in (\ref{eq:landauginz}) one still obtains the $k^{-5/3}$ scaling in the inertial range.  However, the energy spectrum defined with $\bm v$ is significantly
modified deep in the UV for $k > \Lambda_+$ because $\bm v$ is an analytic function of $\bm x$ near vortex cores.}

\section{Energy cascade and Dissipation mechanism} \label{sec:diss}

With Kolmogorov scaling established, we now demonstrate that the system exhibits a direct energy cascade.  
We first establish that dissipation happens exclusively in the UV with the {dominant} dissipation mechanisms being 
vortex drag and vortex annihilation.

As discussed in Section~\ref{sec:set}, a precise measure of dissipation in our system comes from the dual gravitational physics.  In the dual gravitational 
{description,} any energy that flows into the horizon is irreversibly lost and therefore should be thought of as energy lost to heat.  
As Fig.~\ref{fig:candy} suggests, one can analyze how the flux of {energy, $-\tau^{z}_{\ t}$, through} the horizon correlates with location of vortices in the 
superfluid and with vortex annihilation events and thereby assess the dissipation mechanisms.  

\begin{figure}[h!]
\includegraphics[scale = 0.50]{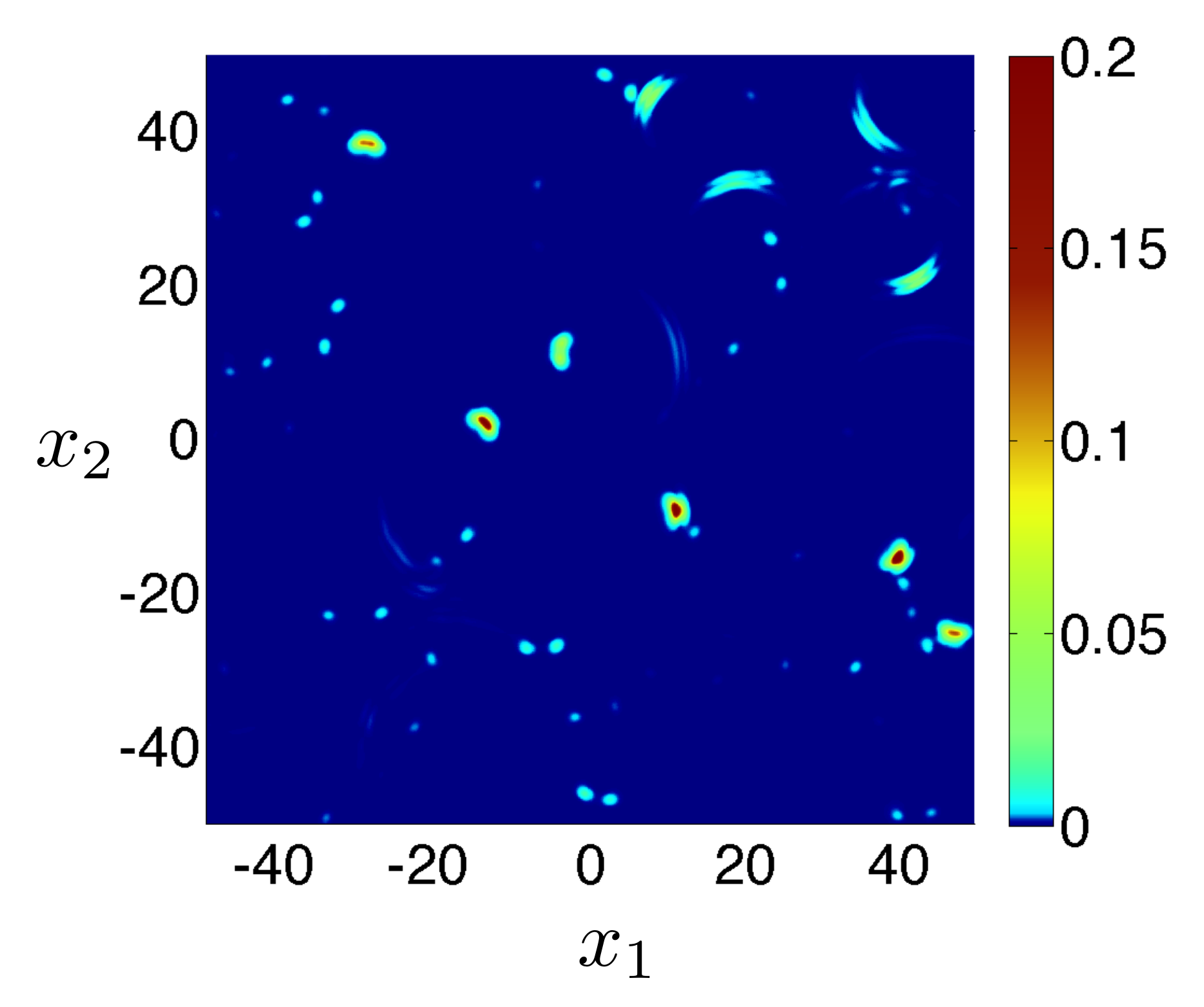}
\vskip -0.05in
\caption{The flux of energy, $-\tau^{z}_{\ t}$, through the horizon at time $t = 300$.  The flux is zero nearly everywhere except at 
the location of vortices (shown in Fig.~\ref{fig:condensate}).  This adds considerable support to the physical picture described in Sec.~\ref{sec:set} and Fig.~\ref{fig:candy}.  The flux is largest during vortex annihilation events.
\label{fig:flux}
} 
\end{figure}

Fig.~\ref{fig:flux}  shows the flux through the horizon at time $t = 300$, the same time as shown in the right panel of Fig.~\ref{fig:condensate}.
The flux is zero nearly everywhere except in the neighborhood of a few isolated points.  Comparing 
the right panel of Fig.~\ref{fig:condensate} to Fig.~\ref{fig:flux}, we see the flux is non-zero at points 
corresponding to the location of vortices.  This contribution to the flux persists at all times and is always 
localized at the position of the vortices.  We therefore identity this contribution to the flux as vortex drag.  Also present in Fig.~\ref{fig:flux} 
are large (but sparse) contributions to the flux from vortex annihilation events.  Again, comparing 
Fig.~\ref{fig:flux} to the right panel of Fig.~\ref{fig:condensate} we see that the flux is largest at the location of vortex pairs in the process 
of annihilating.  Note that the arcs seen in the upper right corner of Fig.~\ref{fig:flux} are remnants of {previous} vortex annihilation events.
The fact that the flux though the horizon is localized at the position of 
vortices adds considerable support to the physical picture described in Sec.~\ref{sec:set} and Fig.~\ref{fig:candy}.

\begin{figure}[h!]
\includegraphics[scale = 0.30]{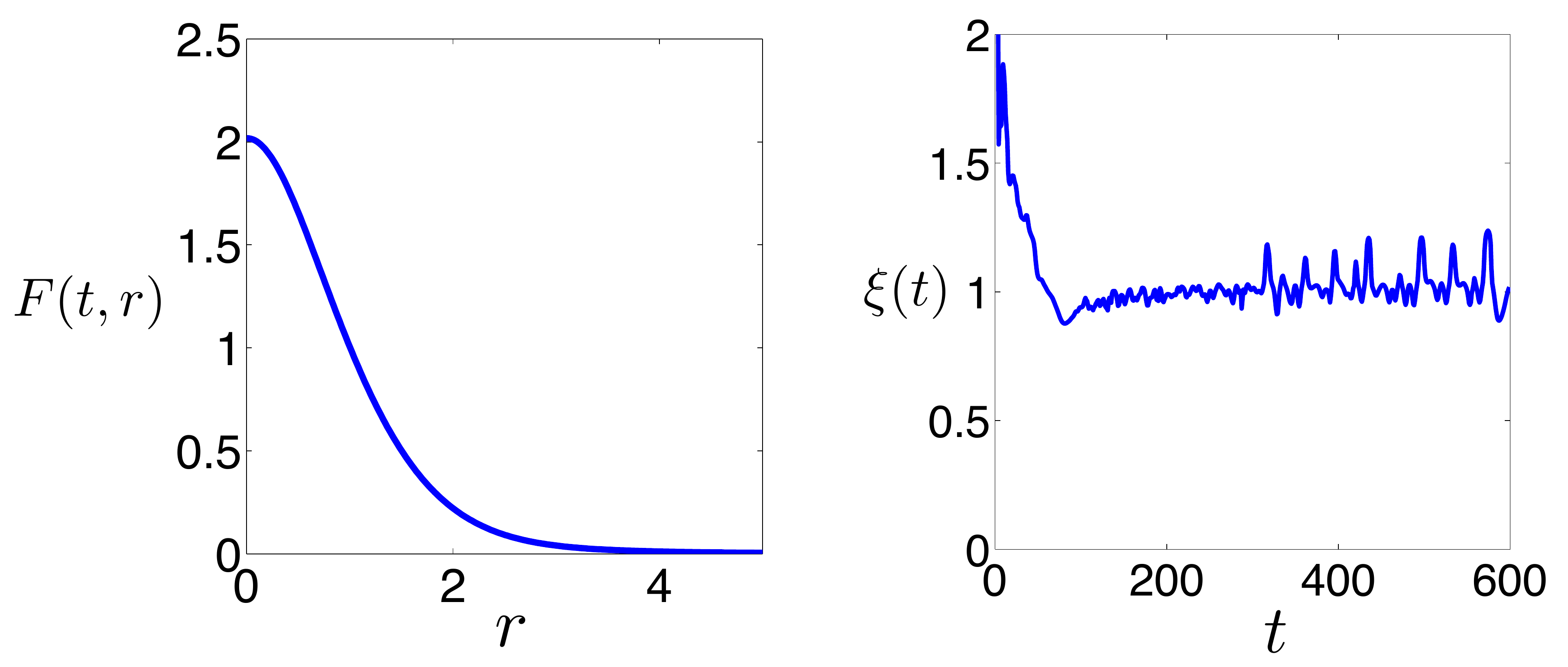}
\vskip -0.05in
\caption{Left: the horizon flux correlation function at time $t = 300$.  Right: the flux correlation length $\xi(t)$ as a function of time.
\label{fig:fluxcorr}
}
\end{figure}

The fact that the energy flux through the horizon is non-zero only in the neighborhood of vortices or vortex annihilation events demonstrates
that energy is dissipated in the UV.  To quantify this statement, in Fig.~\ref{fig:fluxcorr} we plot the flux correlation function
\be 
F (t,r) \equiv \int d^2 x \int d\theta \, \tau^{z}_{\ t}  (t, \bm x + \bm r,z) \, \tau^{z}_{\ t}  (t,\bm x,z) \big |_{\rm horizon},
\ee
at time $t = 300$.  Here $\th$ is the polar angle for $\bm r$ and $r = | \bm r|$ is its norm.  The correlation function is localized 
about $r = 0$ and rapidly vanishes for large $r$.  Also plotted in Fig.~\ref{fig:fluxcorr} is the dissipative correlation length $\xi(t)$ defined by the 
full width half maximum of $F(t,r)$.  After time $t = 100$ $\xi(t) \approx 1$.  {The fact that $\xi(t)$ is roughly constant
reflects the fact that vortex drag and annihilation don't dissipate energy at wildly different scales and that annihilation events, which dissipate at slightly larger length scales than drag, are rare.}  One can therefore define a 
dissipative momentum scale $k_{\rm diss} = 2 \pi/\xi \approx 2 \pi$.  We note that $k_{\rm diss} > \Lambda_+$, so $k_{\rm diss}$ lies 
outside of the inertial range.  

Importantly, the dissipation scales $\xi$ and $k_{\rm diss}$ are controlled by the chemical potential.  For example, repeating the same analysis for turbulent flows with $\mu=7$ modifies the above results in two correlated ways.  First, the mean dissipation correlation length $\xi$ decreases by a factor of roughly 1.3.  Second, the UV knee in the energy spectrum, $\Lambda_+$, which defines the UV end of the scaling regime, increases by a factor of roughly 1.3.  This correlation reinforces the idea that the knee is set by dissipation at the vortex scale and the vortex core size.  

The preference for energy to dissipate in the UV suggests the system is undergoing a direct cascade: 
energy is being transported from the IR through the inertial range $k \in (\Lambda_-,\Lambda_+)$ and dissipated
at $k_{\rm diss} > \Lambda_{+}$.  To test whether this picture is correct we preform the following experiment.
During the scaling regime, we gently drive the system by turning on a weak source for the conserved current $j^\mu$ {and inject energy into the system}. We do this at specific scales {$k_{\rm inject}$}
and examine whether the injected energy gets transferred to other scales {both in the dual gravitational description and in the superfluid description}. 
The crucial tools are again the energy flux through the horizon~\eqref{eq:fl} and the superfluid kinetic energy spectrum~\eqref{eq:ens}.

{In the presence of an external source $a_\mu$ for $j^\mu$, Eq.~(\ref{eq:cons}) can be integrated to give the rate of change of the bulk energy}  
\be 
\p_t \int d^{2} {\bm x} dz \, [- \tau^t{_t}(t,\bm x,z)] =  Q_{\rm boundary} (t)- Q_{\rm horizon} (t)
\ee
where $Q_{\rm horizon}$ was introduced in~\eqref{eq:fl} and $Q_{\rm boundary}$ is the power injected from the boundary 
\be \label{eq:boudf}
Q_{\rm boundary} = -\int d^2 x \, \tau^z{_t} (t,\bm x, z) \big |_{\rm boundary}  =  \ha \int d^2 x \, E_i(t,\bm x) \langle j^i(t,\bm x) \rangle  \ .
\ee
$E_i$ is the boundary ``electric field'' defined by $E_i = \p_t a_i - \p_i a_t$. Up to the prefactor the last equality is of course what one would expect from  electromagnetism.  See Appendix~\ref{app:A} for a derivation of~\eqref{eq:boudf}.  By comparing the controllable injection scale {$k_{\rm inject}$} of $Q_{\rm boundary}$ and measuring the dissipative scale {$k_{\rm diss}$} at the horizon, we can then extract the direction of energy transfer in the dual gravitational description.  

Let us first consider driving the system at long wavelength { $k_{\rm inject} = 0$}. 
During the scaling regime of the turbulent flow discussed in last section we turn on the following homogenous ``electric'' field for a brief period of time
\be \label{eq:sou}
E_x(t) = \eta (t - t_o) g(t - t_o), \qquad E_y(t) = -E_x(t),
\ee
where $\eta$ is a small constant and $t_o = 230$ and $g(t)$ is a gaussian of width $8$.  The electric field first pushes and then pulls so the net momentum transferred to the system is approximately zero and is sufficiently weak so no new vortices are formed.

{In the dual gravitational description, {energy is injected by the electric field} in a $k_{\rm inject}=0$ mode from the boundary ($z = 0$ in Fig.~\ref{fig:candy}).  The injected energy is then transferred via the bulk dynamics to the horizon ($z = 1$ in Fig.~\ref{fig:candy}), where it can dissipate.   In our simulations, the resulting} dissipative correlation length is 
essentially identical to that shown in Fig.~\ref{fig:fluxcorr}.  
Therefore, {the injected} energy is dissipated at the horizon at the scale $k_{\rm diss} = 2 \pi/\xi \sim 2 \pi$.  This transfer of energy --- from the IR at the boundary to the UV at the horizon --- is a 
telltale signature of a direct cascade in the dual gravitational description.

\begin{figure}[ht!]
\includegraphics[scale = 0.45]{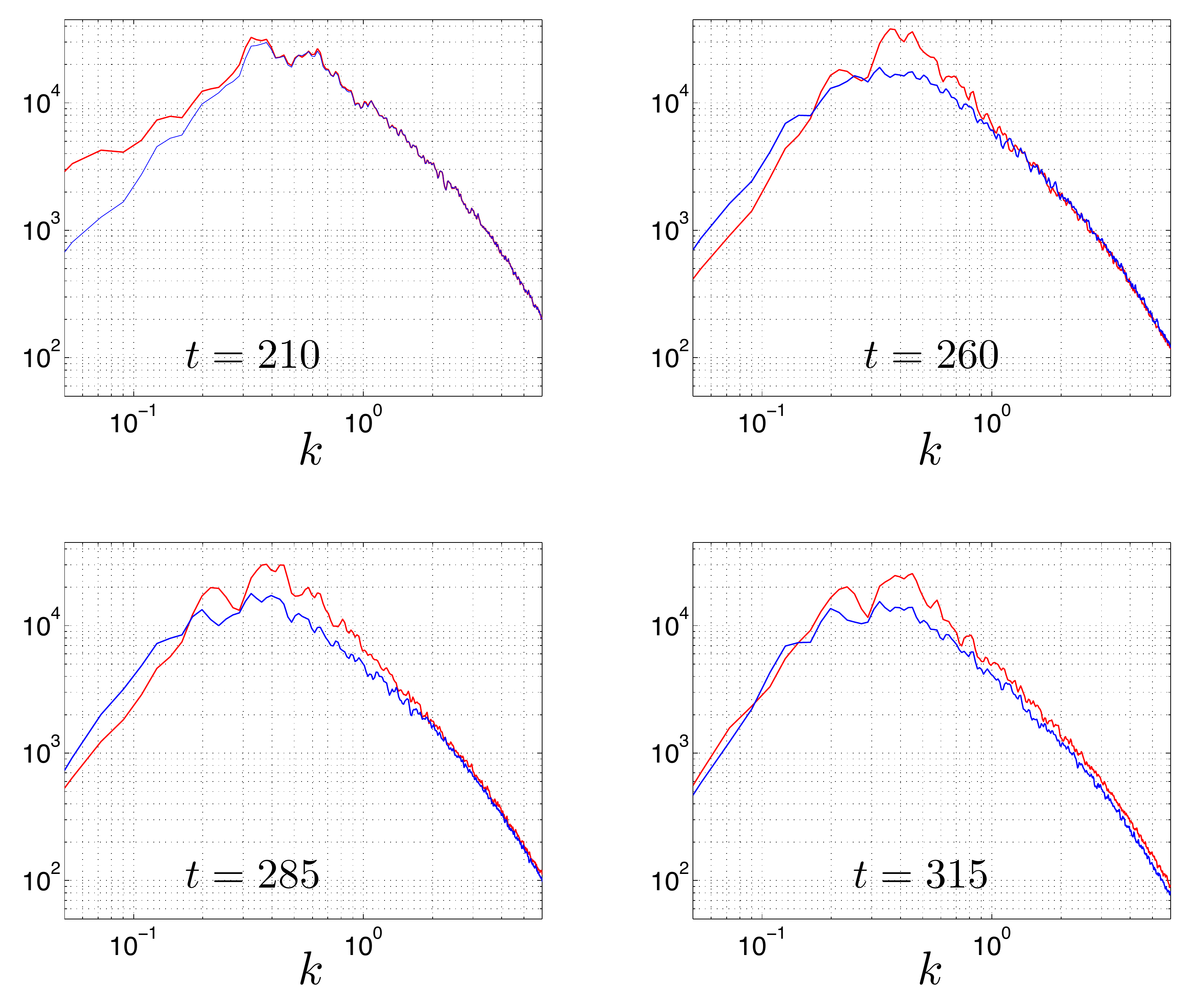}
\vskip -0.05in
\caption{ {The time evolution of the energy spectra for driven and undriven systems.
The blue curve is the energy spectrum with no driving while red curve is that with 
with driving in the IR.  Drive adds energy in the IR between times $ 210 < t < 260$.  
As time progresses the added energy propagates from the IR to the UV where it is dissipated.}
\label{fig:enev}
} 
\end{figure}

One need not rely only on the dual gravitational physics to see that the system is undergoing a direct cascade into the UV.  
One can also see from the evolution of the kinetic energy energy spectrum (\ref{eq:ens}) that the system is undergoing a direct cascade.
Fig.~\ref{fig:enev} shows a comparison of the evolution of the energy spectrum between the driven and undriven systems with the same initial conditions.
When the electric field begins to turn on around time $t = 210$, it adds energy to the system at low $k$.  
When the electric field turns off around time $t = 260$, the {strectra} of the driven and undriven systems agree deep in the UV.  However, there is a significant surplus of kinetic energy 
around $k = 0.4$ for the driven system.  As time progresses, this surplus of energy \textit{propagates} deeper into the UV: there is a flow of energy from the IR to the UV.  
At time $t = 315$ this flow of energy to the UV results in an upwards shift of the entire spectrum for $k > 0.4$ relative to the undriven system.  Again, this behavior is a telltale signature
of a direct cascade.
 
{By contrast, when we inject energy in the UV,  $k_{\rm inject} > \Lam_+$, the injected energy dissipates away without modifying the kinetic energy spectrum in the IR.}

\section{Discussion and outlook}

In this paper we numerically constructed turbulent flows in a $(2+1)$ dimensional holographic superfluid.
These flows exhibited an inertial regime characterized by Kolmogorov scaling, with dissipation dominated by vortex annihilation and vortex drag.
By driving the system in the UV and in the IR, we demonstrated that the observed turbulent behavior involves a direct energy cascade across the inertial regime.  
The gravity description also provides a strikingly simple and intuitive picture for understanding the dissipation mechanism and dissipation scale of the system,
as discussed in Sec.~\ref{sec:set} and depicted in Fig.~\ref{fig:candy}. 

{While our results were obtained at finite temperature $T$, we believe the qualitative physics presented in this paper, and in particular the direction of the cascade, remains the same in low temperature limit.  
This is natural from the perspective of the dual gravitational physics.  In the gravitational description finite temperature is encoded by the presence of black hole whose distance from the boundary of AdS is inversely proportional 
to the temperature.  Therefore, taking the low $T$ limit corresponds to taking the limit that the horizon is very far from the boundary.  However, as argued in Sec.~\ref{sec:set} and as illustrated in Fig.~\ref{fig:candy}, the horizon is screened from $U(1)$ excitations by the presence of a slab of charged condensate.  
The distance from the boundary to the slab is set by the chemical potential.  
The slab effectively decouples dynamics near the boundary from dynamics below the slab.  Consequently, we do not expect a qualitative change in the near-boundary physics (and hence superfluid physics) in the $T \to 0$ limit.}

A question which has been much  discussed in the literature (see~\cite{VivenR,Paoletti2011,Tsubota2012}) is whether the  Kolmogorov scaling observed
in {non-counterflow quantum} turbulence~\cite{Maurer98,smith93,stalp99,Henn2009PRL,Seman2011,nore97,araki02,kobayashi05,parker05} has a classical origin. For example, experiments in~\cite{Maurer98,smith93,stalp99}  studied turbulence on scales much larger than the typical vortex spacing; scales on which one expects superfluid flows to resemble those of classical fluids. 
Our result suggest that quantum effects are crucial, as classical turbulence has an inverse cascade in $(2+1)$ dimensions while quantum turbulence, at least in the systems and regimes we have studied, gives a direct cascade. 
Furthermore, as discussed earlier, in our system average vortex spacing (which is approximate $10$) falls inside the inertial range $(2, 16)$. Since vortex spacing provides the characteristic length scale at which quantum effects are important, the Kolmogorov scaling observed here would appear to be tightly intertwined with the quantum nature of the fluid.  
We should emphasize that the Kolmogorov scaling only assumes the existence of an inertial range {of $k$'s in which the only scale in the system is the overall rate of dissipation, $\varepsilon$. This assumption by itself does not require the fluid to be} classical or quantum, so the Kolmogorov scaling could well arise from quantum phenomena as we are seeing here.  

Because normal fluids in two spatial dimensions typically experience an inverse cascade, it will also be interesting to see how the system behaves as we change  the relative weights of the normal and superfluid components.

There are many questions for further investigation, including a better understanding of 
the physics of vortex drag and annihilation,
the dependence of the turbulent phase on parameters such as the mean vortex density,
and the physics governing the IR end of the inertial range. 
It would be very interesting to study other observables of the superfluid flow such as velocity statistics, 
which have yielded tantalizing differences between classical and quantum turbulence~\cite{paoletti2008,white2010}.   

Because normal fluids in two spatial dimensions typically experience an inverse cascade, it will be interesting to see how the system behaves depending on the relative weights of the normal and superfluid components.
For this purpose we believe it is important to go beyond the probe limit used in this paper.  This will allow one to study the 
the nonlinear dynamics associated with the stress tensor, the charged current, and the nonlinear interactions between the stress tensor 
and the charge current.  These interactions may play an important role in the evolution of the normal fluid component.
In the dual gravitational description going beyond the probe limit is tantamount to including the backreaction of the gauge field $A_M$ and scalar field $\Phi$ on the bulk geometry and will require using numerical relativity to determine the evolution of the system.

To conclude, holographic duality offers  a new laboratory and powerful new tools to study quantized vortex dynamics and quantum turbulence.  We expect it to play an important future role in developing our understanding of  these fascinating phenomena.

\acknowledgments

We acknowledge helpful conversations with Luis Lehner, John McGreevy, Dima Pesin and Laurence Yaffe.
AA thanks the Stanford Institute for Theoretical Physics for hospitality during early stages of this work.  We thank
Sean Hartnoll for pointing out an error in a previous version of this paper.
The work of PC is supported by a Pappalardo Fellowship in Physics at MIT. 
The work of HL is partially supported by a Simons Fellowship.
This research was supported in part by the DOE Office of Nuclear Physics under grant \#DE-FG02-94ER40818.

\appendix

\section{Supplementary materials} \label{app:A}

\subsection{Black hole metric and bulk equations of motion}

Following \cite{Gubser:2008px,Hartnoll:2008vx}, we study a charged holographic superfluid with a global $U(1)$ symmetry.  
The dual holographic description consists of gravity in asymptotically 
AdS$_4$ spacetime coupled to a $U(1)$ gauge field $A_M$ and a scalar field $\Phi$ of charge $q$ with action
\begin{equation}
\label{eq:action}
S = \frac{1}{16 \pi G_{\rm N}} \int d^4x \sqrt{-g} \left [ R + \Lambda  +\frac{1}{q^2} \mathcal L_{\rm matter}  \right ],
\end{equation}
where $\Lambda = -3/L^2$, $L$ is the radius of curvature of AdS, and $G_{\rm N}$ is Newton's constant.  The matter lagrangian is 
\begin{equation}
\mathcal L_{\rm matter} = -\frac{1}{4} F_{MN} F^{MN} - |D \Phi|^2 - m^2 |\Phi|^2,
\end{equation}
with $D_M = \d_M- i A_M$,  $\d_M$ the metric covariant derivative. We take $m^2 = -2/L^2$ which corresponds 
to  $\psi$ {having scaling dimension} $\De =2$.  
We work in a probe limit, i.e. $q$ is large, in which the matter fields decouple from gravity. 
The black hole metric in infalling coordinates can be written as,
\begin{equation}
\label{eq:blackbrane}
ds^2 = \frac{L^2}{z^2} \left [ - f(z)\, dt^2 + d{\bm x}^2 - 2 dt \,dz  \right ] \,.
\end{equation}
Here $t$ is time, $\bm x = \{x_1,x_2\}$
are spatial directions, $z$ is the AdS radial coordinate and $f(z) = 1 - (z/z_{\rm h})^3$.  
The AdS boundary lies at $z = 0$ and the horizon at $z = z_{\rm h}$.  
Lines of constant $(t,\bm x)$ correspond to infalling null geodesics.  
The black hole has a Hawking temperature 
$T = {3\over 4\pi} z_{\rm h}$, 
which is also the temperature of the dual boundary theory.   Without loss of generality we pick 
units where $L = 1$ and $z_h = 1$. 

In the probe limit, the  equations of motion are simply
\begin{align}
\label{eq:eqm}
\d_N F^{MN} = J^N, \ (-D^2 + m^2) \Phi = 0,
\end{align} 
where $J^M = i \Phi^*  D^M \Phi - i \Phi D^M \Phi^*$ is the bulk {electric} current.  We work in axial gauge, $A_{z}$=0.
These equations must be augmented with boundary conditions at the horizon and the boundary.  
At the horizon, in our infalling coordinates, physical solutions should be regular, i.e. infalling.  
Near the boundary,  a general solution takes the form 
\be
A_\nu(t,\bm x,z) \!=\! \CA_\nu(t,\bm x) + O(z), \qquad \Phi(t,\bm x,z) \!=\! z\, \varphi(t,\bm x) + O(z^2) \ .
\ee 
$\CA_\nu$ defines a background gauge field for the $U(1)$ current $\CJ^{\nu}$ of the dual theory, with $\varphi$ an external source for the condensate $\psi$. We are interested in a theory at finite chemical potential $\mu$ with external sources $\varphi$ set to zero, i.e. 
\be
\CA_{t}(t,\bm x) = \mu, \qquad \varphi(t,\bm x) = \CA_{i}(t,\bm x)= 0 \ .
\ee
The expectation value of the superfluid condensate is then determined by the subleading asymptotics of $\Phi$,
\be
\langle \psi(t,\bm x)\rangle =  \lim_{\rm z \to 0} {1\over2}\partial_z^2\, \Phi(t,\bm x,z) \ .
\ee
For definiteness we will choose $\mu =6$ in the aforementioned unit for which $\mu/T = 8 \pi$.  

\subsection{Superfluid phase}

\begin{figure}[h!]
\includegraphics[scale = 0.45]{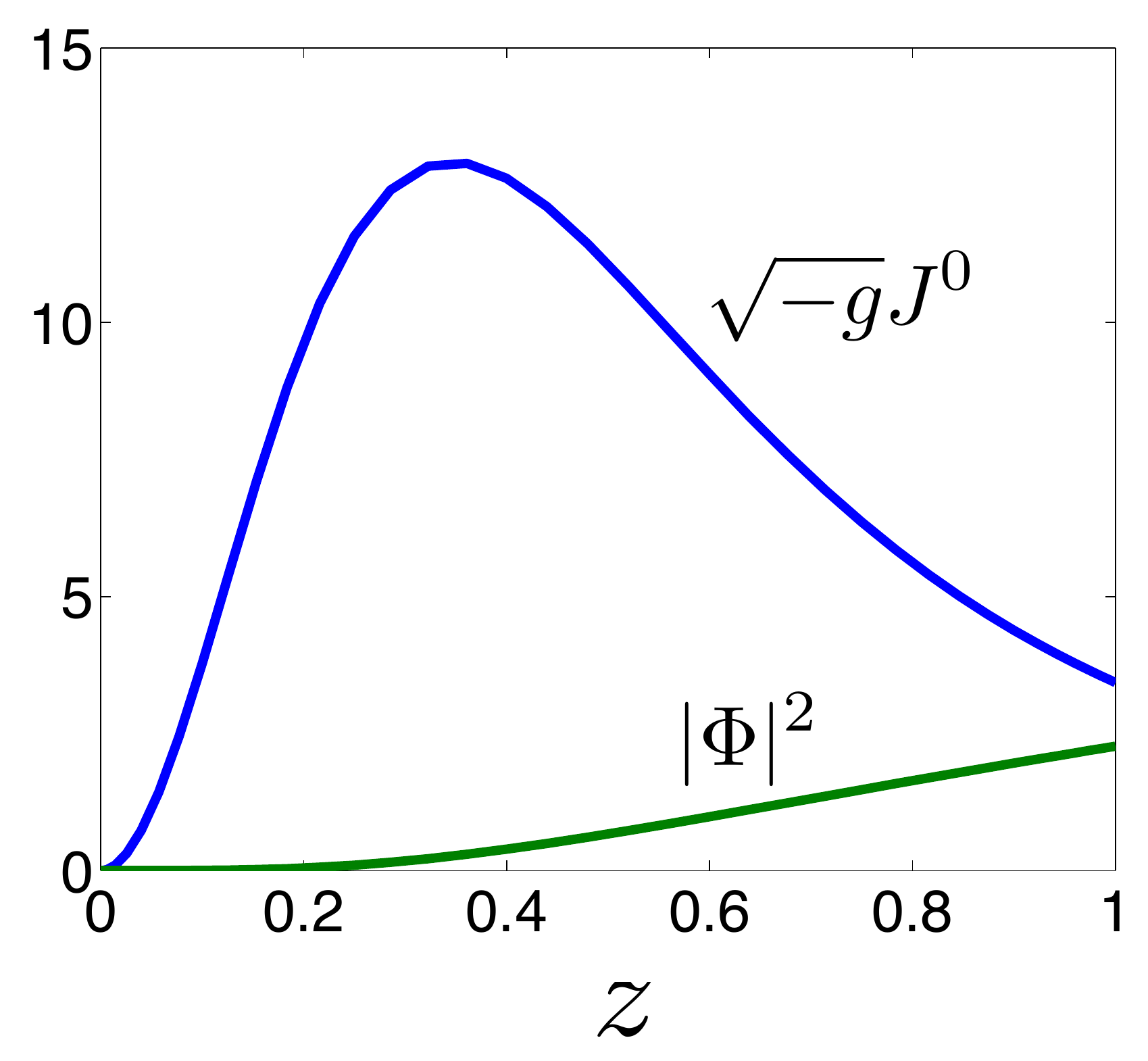}
\vskip -0.05in
\caption{For a uniform boundary condensate, 
the profile of bulk charge density (blue) 
and of the square of
the bulk condensate (green). 
\label{fig:super}
} 
\end{figure}

For a uniform condensate,  
equations~\eqref{eq:eqm} reduce to ordinary differential equations for $\phi (z) \equiv A_t (z)$ and $\Phi (z)$ which can be easily solved, whose solution we denote as $\phi_{\rm eq}$ and $\Phi_{\rm eq}$ respectively. The profile of $|\Phi_{\rm eq} (z)|^2$ and the bulk charge density $\sqrt{-g} J^0 (z)$ are shown in Fig.~\ref{fig:super}.
The charge density distribution has a maximum between the horizon and the boundary, which can be heuristically 
visualized as a charged slab screening the horizon from the boundary excitations.   For the parameters used in this paper, $77\%$ of the total charge lies above the event horizon
and $23\%$ inside the event horizon. 

\subsection{Horizon energy flux} \label{app:hor}

The stress tensor $\mathcal T^{M}_{\ N}$ of the electromagnetic and scalar fields in the bulk is given by
\begin{align}
\nonumber
\mathcal T^{M}_{\ N} = {\textstyle \frac{1}{2}} \big \{  F_{NA}F^{MA} - {\textstyle \frac{1}{4} } \delta^{M}_{\ N} F_{AB}F^{AB} 
+ D_N \Phi^* D^M \Phi \\ \label{eq:bulkstress}
+ D^M \Phi^* D_N \Phi - {\textstyle \frac{1}{2}} \delta^{M}_{\ N}  \left ( D_A \Phi^* D^A \Phi + m^2 \Phi^* \Phi \right ) \big \}.
\end{align}
and satisfies 
\be \label{covco}
d_M \sT^{M}{_ N} =0 \ .
\ee
Now note that for any metric which is independent of $x^\mu$
\be 
\Ga^P_{M\nu}  \sT^M{_P} = 0, \ 
\ee
so\eqref{covco} can be written as a conservation equation
\be
\p_M \tau^M{_\nu} =0 , \qquad \tau^M{_N} \equiv \sqrt{-g} \sT^{M}{_N}  \ .
\ee
We thus have 
\begin{equation}
\partial_\mu \tau^{\mu}_{\ \nu} = -\partial_z \tau^{z}_{\ \nu}\ 
\end{equation}

Now consider turning on external sources $a_\mu$ and $\varphi$ for $j^\mu$ and $\psi$ respectively, i.e. the corresponding bulk field should satisfy the following boundary conditions as $z \to 0$ 
\be 
A_\mu (z, x^\mu)= a_\mu (x^\mu) + \cdots, \qquad 
\Phi = \varphi z^{d - \De}  + \cdots
\ee
 where here $d=3$ and $\De =2$. Introducing 
\be 
E_i = \p_0 a_i - \p_i a_0
\ee
and using the standard AdS/CFT relations 
\be 
\vev{j^\mu} = - \lim_{z \to 0} \sqrt{-g} F^{z \mu} (z), \qquad 
\vev{\psi} =-  \lim_{z \to 0} z^{d - \De} \sqrt{-g} D^z \Phi 
\ee
we find the energy flux near the boundary is given by
\bea 
-\lim_{z \to 0} \tau^z{_0} (z) &= & - \ha  \lim_{z \to 0} \sqrt{-g} \le(  F_{0i}F^{zi}  + D_0 \Phi^* D^z \Phi 
+ D^z \Phi^* D_0 \Phi  \ri) \cr
& =&  \ha E_i \vev{ j^i}  + {\rm Re} \le[ (D_0 \varphi)^* \vev{\psi} \ri] 
\eea
with 
\be 
D_0 \varphi = \p_0 \varphi - i  a_0 \varphi \ .
\ee
For $\varphi=0$ we will then find 
\be 
-\lim_{z \to 0} \tau^z{_0} (z) =  \ha E_i \langle j^i \rangle   \ .
\ee

\subsection{Details on numerical calculation}

In  polar coordinates $\bm x = \{r, \theta\}$ the ansatz for a single vortex solution with winding number $W$ can be written as 
\be 
\Phi_{W} (\bm x,z) = g(r, z) e^{i W \theta}, \qquad A_\th (\bm x,z) = \chi (r,z), \qquad A_t = \phi (r,z) 
\ee
where 
\be 
g(r\to 0, z) \to 0 , \qquad  g (r \to \infty , z) \to \Phi_{\rm eq}(z) \ .
\ee

To construct the scalar field corresponding to a lattice of vortices as in the initial condition discussed in 
Sec.~\ref{sec:tub} we take  $\Phi(t{=}0,\bm x,z) = \Phi_{\rm lat}(\bm x,z) e^{i \chi(\bm x)}$ where
\begin{equation}\label{eq:ic}
\Phi_{\rm lat} (\bm x,z) = 
 \Phi_{\rm eq}(z)
+ \sum_{m,n}  \delta \Phi_{W_{m\!n}}(\bm x {+} m b \hat x_1 {+} n b \bm \hat x_2,z),
\end{equation}
with $b$ the lattice constant and 
$\delta \Phi_{W}(\bm x,z) \equiv \Phi_W(\bm x,z) - \Phi_{\rm eq}(z) e^{i W \theta}$.
{We choose $\chi(\bm x) = {\rm Re} \sum_{\Lambda \leq |\bm k|  \leq 3 \sqrt{2} \Lambda} \alpha (\bm k) e^{i \bm k \cdot \bm x}$ where  $\Lambda  = 2 \pi/100$ and $\alpha(\bm k)$ is a set of $O(1)$ random coefficients. }
We set $\bm A(t{=}0,\bm x,z) = 0$.  In our chosen gauge, 
$A_z = 0$, $A_t$ can be determined from $\bm A$ and $\Phi$ by the constraint equation.

We numerically solve the bulk 
equations of motion 
using pseudospectral methods, 
expanding all functions in a basis of 28 Chebyshev
polynomials in the radial direction and 351 plane waves in each boundary spatial direction.

\bibliography{refs}%
\end{document}